\documentclass{ws-ijmpa}

\begin{document}

\markboth{Deshan Yang} {Recent Development of QCD Factorization for
$B\to M_1 M_2$}

\catchline{}{}{}{}{}
%
%%%%%%%%%%%%%%%%%%%%%%%%%%%%%%%%%%%%%%%%%%%%%%%%%%%%%%%%%%%%%%%%%%%%

\title{RECENT PROGRESS IN QCD FACTORIZATION FOR
$B\to M_1 M_2$\footnote{Invited plenary talk given at the 4th
International Conference on Flavor Physics, Sep.24-Sep.28, 2007,
Beijing. } }

\author{DESHAN YANG}

\address{College of Physical Sciences, the Graduate University of the Chinese
Academy of Sciences, \\Beijing 100049, P.R.China}

\maketitle

\begin{history}
\received{December 28, 2007}
%\revised{Day Month Year}
\end{history}

\begin{abstract}
\noindent After briefly introducing the framework of QCD
factorization for $B\to M_1 M_2$ in the language of the
Soft-Collinear Effective Theory, we firstly address the recent
efforts on higher-order radiative corrections in QCD factorization.
Then we discuss some phenomenologies in $B\to VV$ within the
framework of QCD factorization.

\end{abstract}

\ccode{PACS numbers: 13.20.He,12.60.-i}

\section{Introduction}
$B$ non-leptonic two-body decays provide an abundant sources of
information about the CKM matrix elements as well as possible new
physics effects. However, the complicated strong interaction makes
it difficult to extract the CKM parameters or identify the new
physics signals from the experimental data straightforwardly. Thus,
the more we know about the QCD effects in these decays, the better
we understand the CP violations and new physics.

With the help of the weak effective Hamiltonian, the essential task
to calculate the decay amplitude of $B$ decays is left to
computation of the matrix elements of the effective operators $Q_i$.

Beneke et al proposed a factorization formula for $\langle M_1
M_2\vert Q_i\vert \bar B\rangle$ in the limit $m_b\to \infty$ based
on Feynman diagrams expansion in $\Lambda_{\rm
QCD}/m_b$\cite{Beneke:1999br}. This is so-called QCD factorization
method (also known as BBNS method) for $B\to M_1 M_2$. Here we
present this factorization formula in the language of the
Soft-Collinear Effective Theory (SCET) which is also one of recent
major theoretical development of $B$ physics \cite{Bauer:2000yr}
\begin{eqnarray}
\label{QCDF}
 &&\langle M_{1} M_{2}|Q_i|\bar B\rangle
   =  F^{B\to M_1}\,C_{i}^{I} * f_{M_2}\Phi_{M_2}\nonumber\\
   &
    +& C_i^{II} * J*f_B\Phi_B * f_{M_1}\Phi_{M_1} *
    f_{M_2}^h\Phi_{M_2}^h+{\cal O}(\Lambda_{\rm QCD}/m_b)
\end{eqnarray}
where the perturbatively calculable hard function $C_i^{I,II}$ and
the jet function $J$ are defined at the hard scale $m_b$ and
hard-collinear scale $\sqrt{m_b \Lambda_{\rm QCD}}$ respectively,
and the form-factors $F^{B\to M}$ and light-cone distribution
amplitudes $\Phi_M$ are non-perturbative quantities. The second term
of (\ref{QCDF}) is from the so-called hard spectator interaction
(HSI) which is missing in the literature before
\cite{Beneke:1999br}.

QCD factorization method has become one of major theoretical tools
for $B$ decays, and already been implemented into many
phenomenological studies\cite{Beneke:2003zv}. Here I would like to
review some recent developments in this field by mainly focusing on
efforts towards the higher-order radiative corrections and
phenomenological studies of the polarization puzzles in $B$ decays
to two light vector mesons. Besides QCD factorization, PQCD is
another common-used method in this community. One can refer to the
C.D. L\"u's paper in this proceeding for the comparison between
these two approaches.

\section{Developments on higher-order radiative corrections}

The branching ratios of $B\to \pi \pi$ and $\pi K$ have been
measured very well. The direct CP asymmetries in these decays are
also observed. However, due to the graphical analysis, the large
enhancement to the ratio $|C/T|$ is needed to meet the data, where
$T$, $C$ stand for the color-allowed and color-suppressed
tree-amplitude respectively. This is strongly against the usual
expectation. It leads to so-called $B\to\pi\pi$ and $\pi K$ puzzles.

In QCD factorization, these color-suppressed tree-amplitudes are
related to the QCD coefficient $\alpha_2(M_1 M_2)$. For
illustration\cite{Beneke:2003zv},
\begin{equation}
\alpha_2(\pi\pi) = 0.17_{\rm LO} - [0.17+0.08i\,]_{V_2} +
[0.46]_{\rm LOHSI}
\end{equation}
The accidental cancelation between the leading order (LO) and
next-to-leading order (NLO) vertex corrections ($V_2$) makes the
hard spectator interaction (HSI) very important. The NLO corrections
to the HSI are recently studied by Beneke, Jager and Yang. In their
papers, the next-to-leading order corrections to the jet function
$J$ and hard coefficient $C^{II}_i$ are calculated in
\cite{Beneke:2005gs} and \cite{Beneke:2005vv,Beneke:2006mk}
respectively. Both of these corrections enhance the color-suppressed
amplitude effectively.

 In Table \ref{table1}, the predictions in QCDF with
NLO HSI in a certain parameter setting ($G_4$) is shown. The
agreement between the prediction and experimental data is very good
except for the direct CP asymmetries $A_{CP}$. It means that the
strong phases still need further study. Recently, the efforts
towards the NLO corrections to the imaginary part of the amplitude
has started\cite{Bell:2007tv}.
\begin{table}[t]
\tbl{Predictions in QCDF with NLO HSI vs. experimental data.}{
\begin{tabular}{c|c|c|c|c|c}\hline
${\rm Br}\times 10^6$ & $G_4$ & Exp.
& $A_{CP}$ & $G_4$ & Exp.\\
\hline $\pi^0\pi^-$ & $5.6$ & $5.7\pm 0.4$ & $\pi^0\pi^-$ & $0.00$ &
$0.04\pm 0.05$
\\ \hline
$\pi^+\pi^-$ & $5.7$ & $5.16\pm 0.22$ & $\pi^+\pi^-$ & $0.04$ &
$0.38\pm 0.07$
\\ \hline
$\pi^0\pi^0$ & $0.81$ & $1.31\pm 0.21$ &$\pi^0\pi^0$ & $-0.38$ &
$0.36\pm 0.33$
\\ \hline
$\pi^-\bar{K}^0$ & $22.6$ & $23.1\pm 1.0$ & $\pi^-\bar{K}^0$ &
$0.00$ & $0.008\pm 0.0025$
\\ \hline
$\pi^0 K^-$ & $12.9$ & $12.8\pm 0.6$ & $\pi^0 K^-$ & $-0.05$ &
$0.047\pm 0.026$
\\ \hline
$\pi^+ K^-$ & $20.6$ & $19.4\pm 0.6$ & $\pi^+ K^-$ & $-0.02$ &
$-0.095\pm 0.013$
\\ \hline
$\pi^0\bar{K}^0$ & $9.1$ & $10.0\pm 0.6$ & $\pi^0\bar{K}^0$ & $0.04$
& $-0.12\pm 0.11$
\\ \hline
\end{tabular}\label{table1}}
\end{table}

\section{$B\to VV$ and polarization puzzles}
The fact of the $V-A$ dominance in the Standard Model (SM) shows the
hierarchy of the helicity amplitudes for $B\to VV$
\begin{equation}
{\cal A}_0:{\cal A}_-:{\cal A}_+ = 1 : \frac{\Lambda}{m_b} :
\left(\frac{\Lambda}{m_b}\right)^{\!2} \label{hierarchy}
\end{equation}
with simple estimation by the naive factorization. However,
experimentally, such hierarchy is obeyed in the tree-dominated
decays ($B\to \rho\rho$ and $\omega \rho$), but violated in penguin
dominated $B\to \phi K^*$ system\cite{Aubert:2003mm}. The more
puzzling case happens in $B\to \rho K^*$ system which are
penguin-dominated, in which both transverse polarization is enhanced
in $B^-\to \rho^- \bar K^{*0}$ but suppressed in $\bar B^-\to \rho^0
K^{*-}$.

Analogue to $B\to PP$ and $PV$, we can derive the QCD factorization
formula for each helicity-amplitude of $B\to VV$
\cite{Beneke:2006hg} which obeys the hierarchy in (\ref{hierarchy})
for most of case. The annihilation contribution brings the large
uncertainties to the QCD factorization prediction in $B\to PP$ and
$PV$ as well as $B\to VV$ due to the severe endpoint singularity.
However, the penguin weak annihilation in longitudinal helicity
amplitude is predicted to be small and with small uncertainty,
perhaps due to an accidental cancelation; in negative-helicity
amplitude, the penguin weak annihilation could be very large. This
leads to two consequences immediately:

1) For the tree-dominated decays which are dominated by the
longitudinal polarization, the penguin amplitude gives small
contribution to the branching ratios. As a result, we could have a
better determination of $\sin 2\alpha$ or $\alpha$ from $B\to
\rho\rho$ than $B\to \pi\pi$ and $\pi \rho$. In
\cite{Beneke:2006hg}, the authors get
$\alpha=(85.6^{+7.4}_{-7.3})^\circ$.

2) For the penguin dominated decays, the negative-helicity amplitude
could be large. This could be an solution for the $B\to \phi K^*$
polarization puzzle. However, in this case, QCD factorization loses
its predictive power.

When the final states involve the neutral vectors $\rho^0$, $\omega$
and $\phi$, the electro-magnetic dipole operator will lead to a
violation of (\ref{hierarchy}). This results in an enhancement for
electro-weak penguin in the negative-helicity
amplitude\cite{Beneke:2005we}
\begin{equation}
\Delta P^{\rm EW}(V_1 V_2) \propto  - \frac{2\alpha_{\rm
em}}{3\pi}\,C_{7\gamma,\rm eff} \frac{m_B\bar{m}_b}{m_{V_2}^2}
.\label{dal3ew}
\end{equation}
This effects can be tested in $B\to \rho K^*$ system because
graphically the decay amplitudes for $B\to \rho K^*$ is dominated by
QCD penguin amplitude $P$ and electro-weak penguin amplitude
$P^{EW}$. One has
\begin{eqnarray}
{\cal A}_h( \rho^- \bar K^{*0}):{\cal A}_h( \rho^0 \bar
K^{*-})\simeq P_h:P_h+P^{\rm EW}_h
\end{eqnarray}
where $h$ denotes for the helicities $0,\mp$.

In Table \ref{table3}, the predictions for $B\to \phi K^*$ and $\rho
K^*$ by QCD factorization are listed. ("$\hat\alpha^{c-}_4$ from
data" means the penguin of negative-helicity amplitude is extracted
from the experiments.) One can see that the predictions agree with
the experimental data very well. It reproduces the polarization
pattern of the penguin-dominated $B\to VV$ decays as well.

\begin{table}[t]
\tbl{QCD factorization predictions for
  $B\to \phi K^*$ and $\rho K^*$ vs. experimental data.}
  {\begin{tabular}{lllll}\hline

    \multicolumn{2}{l}{Observable} &
    \multicolumn{2}{l}{Theory} &
    Experiment
    \\\hline

    \multicolumn{2}{l}{}&
    default &
    $\hat{\alpha}_4^{c-}$ from data &
    \\\hline
    ${\rm BrAv} / 10^{-6}$
      & $\phi K^{*-}$
        & $10.1^{+0.5}_{-0.5}{}^{+12.2}_{-7.1}$
        & $10.4^{+0.5}_{-0.5}{}^{+5.2}_{-3.9}$
        & $9.7\pm 1.5$
        \\\hline
      & $\phi\bar{K}^{*0}$
        & $\phantom{1}9.3^{+0.5}_{-0.5}{}^{+11.4}_{-6.5}$
        & $\phantom{1}9.6^{+0.5}_{-0.5}{}^{+4.7}_{-3.6}$
        & $9.5\pm 0.8$
        \\\hline

     &  $\bar{K}^{*0} \rho^-$
        & $5.9^{+0.3}_{-0.3}{}^{+6.9}_{-3.7}$
        & $5.8^{+0.3}_{-0.3}{}^{+3.1}_{-1.9}$
        & $9.2\pm 1.5$
        \\\hline

    &   $K^{*-} \rho^0$
        & $4.5^{+1.5}_{-1.3}{}^{+3.0}_{-1.4}$
        & $4.5^{+1.5}_{-1.3}{}^{+1.8}_{-1.0}$
        & $< 6.1$
        \\\hline

    $f_L / \%$
      & $\phi K^{*-}$
        & $45^{+0}_{-0}{}^{+58}_{-36}$
        & $44^{+0}_{-0}{}^{+23}_{-23}$
        & $50\pm 7$
        \\\hline
      & $\phi\bar{K}^{*0}$
        & $44^{+0}_{-0}{}^{+59}_{-36}$
        & $43^{+0}_{-0}{}^{+23}_{-23}$
        & $49\pm 3$
        \\\hline

     &  $\bar{K}^{*0} \rho^-$
        & $56^{+0}_{-0}{}^{+48}_{-30}$
        & $57^{+0}_{-0}{}^{+21}_{-18}$
        & $48.0\pm 8.0$
        \\\hline
    &   $K^{*-} \rho^0$
        & $84^{+2}_{-3}{}^{+16}_{-25}$
        & $85^{+2}_{-3}{}^{+9}_{-11}$
        & $96^{+6}_{-16}$
        \\\hline
  \end{tabular}
    \label{table3}}
  \end{table}

\section{Acknowledgement}
The author would like to thank the organizers for invitation. The
author is also grateful to M.Beneke and J.Rohrer for collaborations
which the most of this talk is based on. This work is partly
supported by the National Natural Science Foundation of China under
grant number 10705050.


\begin{thebibliography}{99}
%\cite{Beneke:1999br}
\bibitem{Beneke:1999br}
  M.~Beneke, G.~Buchalla, M.~Neubert and C.~T.~Sachrajda,
  %``{QCD} factorization for B --> pi pi decays: Strong phases and CP  violation
  %in the heavy quark limit,''
  Phys.\ Rev.\ Lett.\  {\bf 83} (1999) 1914
  [arXiv:hep-ph/9905312].
  %%CITATION = PRLTA,83,1914;%%

%\cite{Bauer:2000yr}
\bibitem{Bauer:2000yr}
  C.~W.~Bauer, S.~Fleming, D.~Pirjol and I.~W.~Stewart,
  %``An effective field theory for collinear and soft gluons: Heavy to light
  %decays,''
  Phys.\ Rev.\  D {\bf 63} (2001) 114020
  [arXiv:hep-ph/0011336].
  %%CITATION = PHRVA,D63,114020;%%


%\cite{Beneke:2003zv}
\bibitem{Beneke:2003zv}
  M.~Beneke and M.~Neubert,
  %``QCD factorization for B --> P P and B --> P V decays,''
  Nucl.\ Phys.\  B {\bf 675}, 333 (2003)
  [arXiv:hep-ph/0308039], and references therein.
  %%CITATION = NUPHA,B675,333;%%

%\cite{Beneke:2005gs}
\bibitem{Beneke:2005gs}
  M.~Beneke and D.~Yang,
  %``Heavy-to-light B meson form factors at large recoil energy: Spectator
  %scattering corrections,''
  Nucl.\ Phys.\  B {\bf 736}, 34 (2006)
  [arXiv:hep-ph/0508250].
  %%CITATION = NUPHA,B736,34;%%


%\cite{Beneke:2005vv}
\bibitem{Beneke:2005vv}
  M.~Beneke and S.~Jager,
  %``Spectator scattering at NLO in non-leptonic B decays: Tree amplitudes,''
  Nucl.\ Phys.\  B {\bf 751}, 160 (2006)
  [arXiv:hep-ph/0512351].
  %%CITATION = NUPHA,B751,160;%%


%\cite{Beneke:2006mk}
\bibitem{Beneke:2006mk}
  M.~Beneke and S.~Jager,
  %``Spectator scattering at NLO in non-leptonic B decays: Leading penguin
  %amplitudes,''
  Nucl.\ Phys.\  B {\bf 768}, 51 (2007)
  [arXiv:hep-ph/0610322].
  %%CITATION = NUPHA,B768,51;%%


%\cite{Bell:2007tv}
\bibitem{Bell:2007tv}
  G.~Bell,
  %``NNLO Vertex Corrections in charmless hadronic B decays: Imaginary part,''
  arXiv:0705.3127 [hep-ph].
  %%CITATION = ARXIV:0705.3127;%%


%\cite{Aubert:2003mm}
\bibitem{Aubert:2003mm}
  B.~Aubert {\it et al.},  %[BABAR Collaboration],
  %``Rates, polarizations, and asymmetries in charmless vector-vector B  meson
  %decays,''
  Phys.\ Rev.\ Lett.\  {\bf 91}, 171802 (2003);
  %%CITATION = HEP-EX 0307026;%%
%\cite{Aubert:2004xc}
%\bibitem{Aubert:2004xc}
%  B.~Aubert {\it et al.}  [BABAR Collaboration],
  %``Measurement of the B0 $\to$ Phi K0 decay amplitudes,''
  Phys.\ Rev.\ Lett.\  {\bf 93}, 231804 (2004);
  %%CITATION = HEP-EX 0408017;%%
%\cite{Aubert:2004qb}
%\bibitem{Aubert:2004qb}
%  B.~Aubert  [BABAR Collaboration],
  %``Measurements of branching fraction, polarization, and  direct-CP-violating
  %charge asymmetry in B+ $\to$ K*0 rho+ decays,''
  [hep-ex/0408093];
  %%CITATION = HEP-EX 0408093;%%
%\cite{Chen:2003jf}
%\bibitem{Chen:2003jf}
  K.~F.~Chen {\it et al.}, %[Belle Collaboration],
  %``Measurement of branching fractions and polarization in B $\to$ Phi K(*)
  %decays,''
  Phys.\ Rev.\ Lett.\  {\bf 91}, 201801 (2003);
  %%CITATION = HEP-EX 0307014;%%
%\cite{Chen:2005zv}
%\bibitem{Chen:2005zv}
%  K.~F.~Chen {\it et al.}  [BELLE Collaboration],
  %``Measurement of polarization and triple-product correlations in B $\to$  Phi
  %K* decays,''
  Phys.\ Rev.\ Lett.\  {\bf 94}, 221804 (2005);
  %%CITATION = HEP-EX 0503013;%%
%\cite{Zhang:2005iz}
%\bibitem{Zhang:2005iz}
  J.~Zhang {\it et al.},  %[BELLE Collaboration],
  %``Measurements of branching fraction and polarization in B+ $\to$ rho+ K*0
  %decay,''
  [hep-ex/0505039].
  %%CITATION = HEP-EX 0505039;%%


\bibitem{Beneke:2006hg}
  M.~Beneke, J.~Rohrer and D.~Yang,
  %``Branching fractions, polarisation and asymmetries of B -> VV decays,''
  Nucl.\ Phys.\  B {\bf 774}, 64 (2007)
  [arXiv:hep-ph/0612290].
  %%CITATION = NUPHA,B774,64;%%

%\cite{Beneke:2005we}
\bibitem{Beneke:2005we}
  M.~Beneke, J.~Rohrer and D.~Yang,
  %``Enhanced electroweak penguin amplitude in B --> V V decays,''
  Phys.\ Rev.\ Lett.\  {\bf 96}, 141801 (2006)
  [arXiv:hep-ph/0512258].
  %%CITATION = PRLTA,96,141801;%%


\end{thebibliography}
\end{document}